# Above-room-temperature giant thermal conductivity switching in spintronic multilayer

Hiroyasu Nakayama,[1,a] Bin Xu,[2] Sotaro Iwamoto,[2] Kaoru Yamamoto,[1] Ryo Iguchi,[1] Asuka Miura,[1] Takamasa Hirai,[1] Yoshio Miura,[1] Yuya Sakuraba,[1,3] Junichiro Shiomi,[2,a] and Ken-ichi Uchida[1,2,4,5,a]

[1]National Institute for Materials Science, Tsukuba 305-0047, Japan.
[2]Department of Mechanical Engineering, The University of Tokyo, Tokyo 113-8656, Japan.
[3]PRESTO, Japan Science and Technology Agency, Saitama 332-0012, Japan.
[4]Institute for Materials Research, Tohoku University, Sendai 980-8577, Japan.
[5]Center for Spintronics Research Network, Tohoku University, Sendai 980-8577, Japan.

[a]Author to whom correspondence should be addressed:
NAKAYAMA.Hiroyasu@nims.go.jp, shiomi@photon.t.u-tokyo.ac.jp, and UCHIDA.Kenichi@nims.go.jp

## ABSTRACT

Thermal switching provides an effective way for active heat flow control, which has recently attracted increasing attention in terms of nanoscale thermal management technologies. In magnetic and spintronic materials, the thermal conductivity depends on the magnetization configuration: this is the magneto-thermal resistance effect. Here we show that an epitaxial Cu/Co$_{50}$Fe$_{50}$ multilayer film exhibits giant magnetic-field-induced modulation of the cross-plane thermal conductivity. The magneto-thermal resistance ratio for the Cu/Co$_{50}$Fe$_{50}$ multilayer reaches 150% at room temperature, which is much larger than the previous record high. Although the ratio decreases with increasing the temperature, the giant magneto-thermal resistance effect of ~100% still appears up to 400 K. The magnetic field dependence of the thermal conductivity of the Cu/Co$_{50}$Fe$_{50}$ multilayer was observed to be about twice greater than that of the cross-plane electrical conductivity. The observation of the giant magneto-thermal resistance effect clarifies a potential of spintronic multilayers as thermal switching devices.

Thermal switching enables active control of heat transfer, which is a key component of nanoscale thermal management technologies for electronic devices.[1,2] Towards the development of such technologies, thermal switching devices showing large thermal conductivity change, wide temperature range operation above room temperature, high heat exchange, and easy integration into existing electronic devices are desired. In this perspective, various solid-state thermal switching techniques, including metal-insulator transition,[3,4] electrochemical intercalation,[5,6] and electric field control of domain structures[7] and magnons,[8] have been investigated so far. However, development of thermal switching devices exhibiting large thermal conductivity change and wide temperature range operation remains major challenges.

In the emerging field of spin caloritronics, the spin degree of freedom is introduced into thermal and thermoelectric transport phenomena to create novel physics and functionalities.[9,10] In particular, the thermo-spin and magneto-thermoelectric effects in magnetic materials and multilayer structures have been actively investigated since these phenomena offer an unconventional approach to develop energy harvesting and thermoelectric cooling devices with a simple structure, versatile scaling capability, and unique symmetry.[11–15] However, there are only several reports focusing on the active control of thermal conductivity based on spin caloritronics.[16–18]

In magnetic and spintronic materials, the thermal conductivity depends on the magnetization configuration, which is called the magneto-thermal resistance (MTR) effect.[19] The MTR effect makes it possible to achieve not only large magnetic-field-induced change of thermal conductivity but also thermal switching with no-wiring, non-volatile, and wide temperature range operation. Because of these potential advantages, studies on the active control of thermal conductivity in magnetic and spintronic materials are important from the viewpoints of both fundamental physics and thermal management applications.

In this study, we focus on the MTR effect in a spintronic multilayer film comprising alternately-stacked ferromagnetic and nonmagnetic metals and report the observation of the giant magnetic-field-induced switching of the cross-plane thermal conductivity [Fig. 1(a)]. Such a multilayer film is known to exhibit the giant magnetoresistance (GMR).[20] In GMR, the electrical resistivity of the multilayer film is larger when the magnetization directions of the adjacent ferromagnetic layers are aligned antiparallel than when they are parallel; switching of the magnetization configuration can be achieved by a magnetic field or spin injection. GMR is indispensable in data storage and magnetic sensor applications, and has provided an opportunity for fundamental studies of spin-polarized electron transport. The thermal analog of GMR is the MTR effect in spintronic multilayers. A typical system showing MTR is a Cu/Co multilayer film, the in-plane and cross-plane thermal conductivities of which were measured by the $3\omega$ method[16,17] and the time-domain thermoreflectance (TDTR) method,[18] respectively. MTR for the cross-plane thermal conductivity of the Co/[Cu/Co]$_{39}$ multilayer was observed to be greater than that for the in-plane thermal conductivity,[17,18] which is the same tendency as the case in GMR. In the conventional spintronic multilayers, the maximum value of the MTR ratio $(\kappa_\text{P} - \kappa_\text{AP})/\kappa_\text{AP}$ is less than 90%,[16–18] where $\kappa_\text{P}$ and $\kappa_\text{AP}$ are the thermal conductivity for the parallel magnetization configuration (P configuration) and the antiparallel magnetization configuration (AP configuration), respectively. Although these previous studies suggest the usefulness of thermal switching based on MTR, a greater effect is required for its thermal management applications. Recently, Fathoni *et al.*

reported the observation of the very large current-in-plane (CIP)-GMR effect in an epitaxial Cu/CoFe multilayer structure with a metastable bcc-Cu spacer,[21] where the magnetoresistance (MR) ratio $(R_{AP} - R_P)/R_P$ with $R_{P(AP)}$ being the electrical resistance for the P (AP) configuration reaches 73% at room temperature. Therefore, it is interesting to investigate MTR in such a metastable Cu/CoFe multilayer.

Here, we measured the cross-plane thermal conductivity of a similar Cu/Co$_{50}$Fe$_{50}$ multilayer film in the temperature range from room temperature to 400 K and found that the multilayer exhibits giant MTR. Surprisingly, the cross-plane thermal conductivity change $\kappa_P - \kappa_{AP}$ and the MTR ratio reach 24.8 Wm$^{-1}$K$^{-1}$ and 150% at room temperature, respectively; the MTR ratio is the highest value reported so far. The MTR ratio of more than 100% is maintained up to around 400 K. We quantitatively compare the observed MTR effect with the current-perpendicular-to-plane (CPP)-GMR in the same multilayer, and reveal that the MTR ratio is much greater than the MR ratio for the CPP configuration. The giant magnetic-field-induced modulation of the thermal conductivity observed here indicates a great potential of spintronic multilayers as a tool for active thermal management. Since the composition of the multilayer is fixed in this study, hereafter, Co$_{50}$Fe$_{50}$ is referred to as CoFe for simplicity.

To investigate the MTR effect, we measured the cross-plane thermal conductivity of the Cu/CoFe multilayer film by means of the TDTR method based on optical pump-probe techniques.[22,23] In the TDTR measurement, a sample is heated by an ultrafast laser pulse and the transient response of the sample temperature is detected through thermoreflectance [Fig. 1(b)] (see Sec. 1 in supplementary material). The TDTR signals are then discussed with a heat conduction model, enabling the determination of thermal conductivity of thin films. Since the TDTR method does not require any electrical contacts, nanofabrication processes are not necessary for sample preparation procedures.

Figure 2(a) shows a schematic of the stacking structure of the sample used for the TDTR measurements. We deposited the fully-epitaxial CoFe/[Cu/CoFe]$_{33}$ multilayer film with the Cu/CoFe bilayer number of 33 on a MgO (001) substrate (see Sec. 2 in supplementary material). Here, the thickness of the CoFe (Cu) layer is 3.0 (1.6) nm, and the total thickness of the multilayer film is 154.8 nm. The top surface of the multilayer was covered by a thin Al transducer layer with the known thermoreflectance coefficient.[24] The TDTR measurements were performed in the front face heating/front face detection configuration[25] with applying a magnetic field $H$ along the film plane direction [Fig. 1(b)]. Figure 2(b) shows the $H$ dependence of the normalized magnetization $M/M_s$ of the CoFe/[Cu/CoFe]$_{33}$ multilayer film at room temperature, where $M_s$ is the saturation magnetization. The magnetization curve shows almost zero net magnetization at zero field, indicating the AP configuration at $H \sim 0$ kOe. The magnetization increases with increasing $H$ and saturates around $|H| = 0.8$ kOe, indicating the P configuration for $|H| > 0.8$ kOe.

Now, let us see the results of the TDTR measurements at zero magnetic field and room temperature. As shown in Fig. 2(c), the TDTR signals decay in the order of hundreds of picoseconds. This behavior can be understood by transient heat transfer from the surface heating since the TDTR signal is changed in response to the temperature change near the film surface. The effective cross-plane thermal conductivity $\kappa$ of the CoFe/[Cu/CoFe]$_{33}$ multilayer film was obtained using the model that regards the multilayer film as a homogeneous medium (see Sec. 3 in supplementary material). The $\kappa$ value as well as the interfacial thermal conductance $G$ between the bottom CoFe layer and MgO substrate were determined by fitting to the TDTR signals using the heat conduction model. We confirmed that the model has proper sensitivity to $\kappa$ by the sensitivity calculation with nominal and literature values of the thermal transport properties in our system (see Sec. 3 in supplementary material). The fitting results [solid curves in Fig. 2(c)] are in good agreement with the experimental results. We note that $\kappa$ of the Cu/CoFe multilayer film should show no dependence on the Cu/CoFe bilayer number for the thickness range that can be measured by the TDTR method, where the total thickness of the Cu/CoFe multilayer is much larger than the electron/phonon/magnon mean free path.[26-28]

Next, we focus on the TDTR signals under a finite $H$. As shown in Fig. 2(c), the TDTR signals for the CoFe/[Cu/CoFe]$_{33}$ multilayer clearly depend on $H$. From these experimental results and the aforementioned analyses, the $H$ dependences of $\kappa$ of the CoFe/[Cu/CoFe]$_{33}$ multilayer and $G$ at the MgO/CoFe interface were obtained at room temperature [Fig. 2(d)]. We found that the CoFe/[Cu/CoFe]$_{33}$ multilayer exhibits a drastic $\kappa$ change for $|H| < 0.8$ kOe, while $\kappa$ is almost constant for $|H| > 0.8$ kOe. This field range in which the $\kappa$ change appears coincides with the magnetization process of the multilayer, indicating that $\kappa$ shows the maximum (minimum) value for the P (AP) configuration. We also checked that $G$ is almost independent of $H$ [see the inset of Fig. 2(d)], confirming that the $H$ dependence of the TDTR signals originates from the $H$ dependence of $\kappa$, *i.e.*, the MTR effect of the CoFe/[Cu/CoFe]$_{33}$ multilayer. Importantly, our CoFe/[Cu/CoFe]$_{33}$ multilayer film exhibits not only the record-high MTR ratio of $(\kappa_P - \kappa_{AP})/\kappa_{AP} = 150\%$ but also the large cross-plane thermal conductivity change of $\kappa_P - \kappa_{AP} = 24.8$ Wm$^{-1}$K$^{-1}$ at room temperature. The obtained $\kappa_P - \kappa_{AP}$ value is about twice larger than the value for conventional Cu/Co multilayer films[18] and is an order of magnitude larger than the thermal conductivity change due to other thermal switching principles.[3–8]

As shown above, we have observed the giant MTR effect in the Cu/CoFe multilayer film. Next, we show how the giant thermal switching properties change above room temperature. To do this, we performed the TDTR measurements and analyses for the same CoFe/[Cu/CoFe]$_{33}$ multilayer in the $T$ range from room temperature to 400 K, which is much lower than the Curie temperature of the CoFe film.[29] Here, we applied $H = 1.2$ kOe (0 kOe) to measure the cross-plane thermal conductivity for the P (AP) configuration. Figure 3(a) shows the $T$ dependence of the cross-plane thermal conductivity for the CoFe/[Cu/CoFe]$_{33}$ multilayer film, where the red (blue) data points are for the P (AP) configuration. We found that $\kappa_P$ and $\kappa_{AP}$ of the CoFe/[Cu/CoFe]$_{33}$ multilayer film monotonically increase with increasing $T$. The $\kappa_P - \kappa_{AP}$ value also monotonically increases with $T$ and reaches 30.5 Wm$^{-1}$K$^{-1}$ at 400 K [see the inset of Fig. 3(a)]. Figure 3(b) shows the $T$ dependence of the MTR ratio for the CoFe/[Cu/CoFe]$_{33}$ multilayer film. Although the MTR ratio monotonically decreases with increasing $T$, the giant thermal switching with the MTR ratio of ~100% still appears at 400 K.

In Fig. 3(b), we also compare the switching ratio between the thermal and electrical conductivities of the Cu/CoFe multilayer film. To do this, it is important to measure the $T$ dependence of the MR ratio in the CPP configuration since we focus on the cross-plane thermal conductivity, although the conventional study is often based on the comparison between the cross-plane thermal conductivity and in-plane electrical conductivity.[20] We realized the quantitative comparison of the MTR effect with CPP-GMR by patterning the multilayer film into a pillar structure (see Sec. 4 in supplementary material). We found that, in our multilayer film, the MTR ratio is much greater than the MR ratio in the CPP configuration over the $T$ range of interest. This is one of the important behaviors for discussing the giant MTR effect in the Cu/CoFe multilayer film.

Now, we are in a position to discuss the origin of the observed giant MTR effect in the Cu/CoFe multilayer film. Although the MTR effect in spintronic multilayers has been discussed in terms of the spin heat accumulation and spin-dependent temperatures,[30] the phenomenological formulation predicts that the MTR ratio is smaller than the MR ratio because of the short relaxation time for the spin-dependent temperatures. The contribution of the phonon thermal conductivity that is believed to be spin-independent further reduces the MTR ratio. Therefore, the MTR ratio greater than the MR ratio in our Cu/CoFe multilayer film is beyond the conventional expectations based on simple electronic transport. In contrast, it is known that the linear relationship between the electrical and thermal conductivities, *i.e.*, the Wiedemann-Franz law, is violated when the density of states shows a steep change near the Fermi energy.[31] To check the contribution of such electronic effects, we perform the first-principles calculations of the MTR effect with a particular focus on the spin-dependent electron transport at the Cu/CoFe interfaces (note that the importance of the Cu/CoFe interfaces in the MTR effect is suggested by the fact that the thickness of each layer is comparable to or smaller than the electron mean free path and $\kappa$ of our multilayer film is much smaller than the thermal conductivities of Cu and CoFe slabs at room temperature, which were observed to be 425 and 144 Wm$^{-1}$K$^{-1}$, respectively). As shown in Sec. 5 in supplementary material, our first-principles calculation clarifies that the MTR ratio can be greater than the MR ratio depending on the electronic structure and temperature. However, the calculated difference between the MTR and MR ratios is too small to explain the observed giant MTR. We also confirmed that the correction factor for the electronic thermal conductivity due to the thermoelectric effect, *i.e.*, $\sigma S^2 T$ with $\sigma$ and $S$ respectively being the electrical conductivity and the Seebeck coefficient,[32] is approximately 1 Wm$^{-1}$K$^{-1}$ in our Cu/CoFe multilayer film, which cannot explain the experimental results. These discussions suggest that not only conduction electrons but also other heat carriers may contribute to the MTR effect in the spintronic multilayer systems.

One of the possible additional carriers that contribute to the MTR effect is the magnon, collective dynamics of localized magnetic moments in magnetic materials. The magnon thermal conductivity of magnetic materials is typically measured only at low temperatures. However, recently reported simulations on the spin-lattice dynamics suggest that the magnon contribution in the total thermal conductivity can be substantial even in ferromagnetic metals at room temperature.[33] Since CoFe is known to have very small Gilbert damping,[34] it may be a good medium for magnon thermal transport. Here, let us recall the work on the spin Seebeck effect in magnetic multilayers;[35] this study shows that magnon transport in alternately-stacked nonmagnet/ferromagnet multilayers can be modulated via the interaction with conduction-electron spin currents at the interfaces. More recently, the magnetization-configuration-dependent magnon transport called the magnon valve effect is demonstrated in magnetic multilayer structures, indicating the importance of the magnon-magnon interaction between adjacent magnetic layers.[36,37] We thus expect that the magnon thermal conductivity in spintronic multilayers can be switched depending on the magnetization configuration via boundary condition changes in the spin sector. Here, we would like to emphasize the fact that the metastable bcc-Cu spacer used here has a perfect lattice matching at the bcc-Cu/CoFe interfaces;[21] our Cu/CoFe multilayer film is thus suitable for long-range propagation of magnons and electron spins, and may exhibit large magnon-mediated thermal conduction. Although we performed the MTR measurements only above room temperature, low temperature measurements are important to demonstrate this speculation. In the low temperature region where the thermal excitation of magnons can be sufficiently suppressed, the electron contribution to MTR becomes dominant, and the MTR ratio is expected to approach the CPP-GMR ratio. The MTR measurements under strong magnetic fields will also be useful to separate the electron and magnon contributions.[38] To clarify the magnon contribution in MTR, the investigation on the magnon-phonon interaction is also necessary.[39] These systematic measurements will provide a microscopic understanding of the giant MTR effect.

Finally, we discuss future prospects of the giant thermal conductivity switching in spintronic multilayers from a technical point of view. In this study, we have demonstrated the magnetic-field-induced switching of the thermal conductivity with no hysteresis behaviors, where the high (low) thermal conductivity state is maintained only under a finite (zero) magnetic field. However, non-volatile and highly-controllable thermal switching is preferred for thermal management applications. Non-volatile thermal switching based on the MTR effect may be possible by introducing spin-valve structures with exchange bias,[40] which makes it possible to select and maintain a high or low thermal conductivity state even in the absence of external magnetic fields owing to the exchange-bias-induced hysteresis in magnetization processes. Exploring materials and developing multilayer devices showing larger MTR ratio are also necessary; one of the strategies to realize larger MTR is to utilize magnetic tunnel junctions showing a large tunnel magnetoresistance effect, since its MR ratio is typically larger than that for CPP-GMR.[41] The MTR effect in magnetic tunnel junctions remains to be observed, while other spin caloritronic phenomena, such as the magneto-Seebeck and Peltier effects,[42-44] have already been observed. As discussed above, magnon engineering for thermal transport will be important also from the viewpoint of thermal management applications. In light of the fact that the Cu/CoFe multilayer film used in this study was originally designed for CIP-GMR, it is worth investigating the MTR effect for an in-plane thermal conductivity, while this study focuses on that for the cross-plane thermal conductivity.

In summary, we have demonstrated the giant thermal conductivity switching by utilizing simple spintronic multilayer comprising Cu and CoFe films. We obtained the large thermal conductivity change of 24.8 (30.5) $Wm^{-1}K^{-1}$ due to the MTR effect at room temperature (400 K), which is an order of magnitude greater than typical values obtained by other thermal switching techniques.[3–8] The MTR ratio for our Co/CoFe multilayer film reaches 150% at room temperature, which is the highest value reported so far, and the ratio of more than 100% is kept up to around 400 K. We also found that the MTR ratio is about twice greater than the MR ratio for CPP-GMR in the Co/CoFe multilayer film. The observed MTR effect is too large to be explained only by electronic thermal transport, implying a potential contribution of magnons in the thermal conductivity of spintronic multilayers. These findings pave the way for active thermal management technologies for electronic and spintronic devices.

See the supplementary material for more details on TDTR measurements, sample preparation, heat conduction models for Cu/CoFe multilayer, CPP-GMR effect, and first principles calculations of the MTR effect.

The authors thank G. E. W. Bauer and K. Hono for valuable discussions and K. B. Fathoni, W. Zhou, Y. Fujita, N. Kojima, and B. Masaoka for technical supports. This work was supported by CREST "Creation of Innovative Core Technologies for Nano-enabled Thermal Management" (JPMJCR17I1) from JST, Japan, and Grant-in-Aid for Scientific Research (S) (JP17H06152, JP18H05246) and Grant-in-Aid for Scientific Research (A) (JP19H00744) from JSPS KAKENHI, Japan. H.N. would like to acknowledge the ICYS Research Fellowship, NIMS, Japan. A.M. and T.H. are supported by JSPS through Research Fellowship for Young Scientists (JP18J02115, JP20J00365).
## DATA AVAILABILITY

The data that support the findings of this study are available from the corresponding author upon reasonable request.

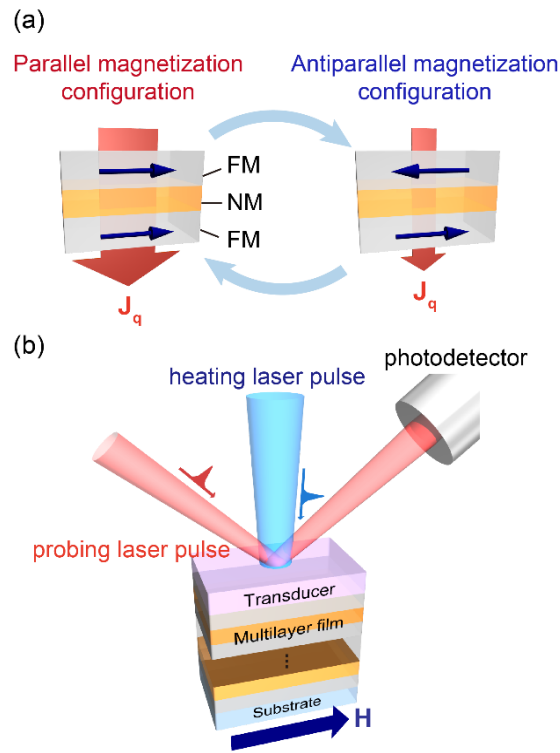

**Fig. 1.** Thermal conductivity switching in spintronic multilayer. (a) Schematic of the magneto-thermal resistance (MTR) effect in a multilayer film comprising alternately-stacked ferromagnetic metal (FM) and nonmagnetic metal (NM) layers. $\mathbf{J}_q$ denotes a heat current. Blue arrows represent the magnetization direction of the FM layers. The thermal conductivity of the multilayer film is switched by changing the magnetization configuration of the FM layers from parallel to antiparallel, and vice versa. (b) Schematic of the time-domain thermoreflectance (TDTR) measurement. **H** denotes an external magnetic field.

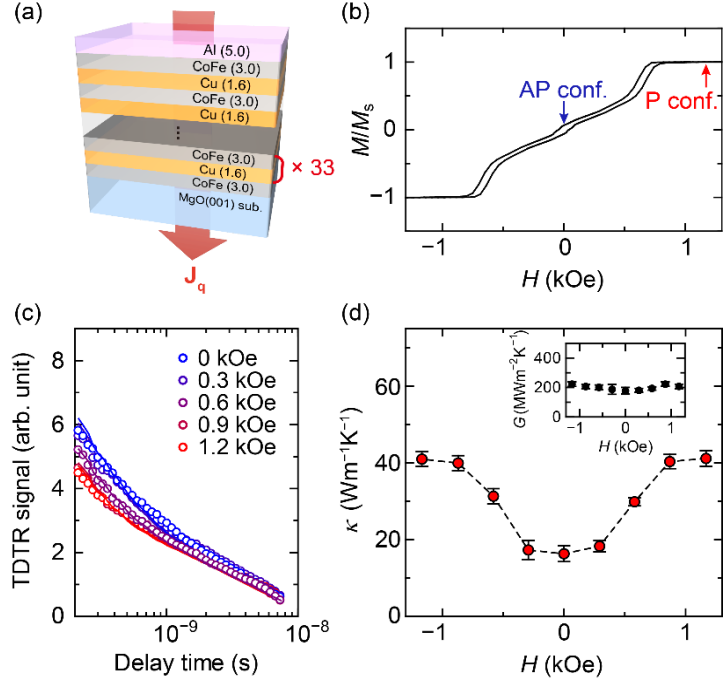

**Fig. 2.** Observation of giant magneto-thermal resistance effect. (a) Schematic of the CoFe/[Cu/CoFe]$_{33}$ multilayer film used for the TDTR measurements. (b) $H$ dependence of the normalized magnetization $M/M_s$ for the CoFe/[Cu/CoFe]$_{33}$ multilayer film at room temperature. (c) Temporal response of TDTR signals for the CoFe/[Cu/CoFe]$_{33}$ multilayer film at room temperature for various values of $H$. The solid curves are the best fits to the data. (d) $H$ dependence of the cross-plane thermal conductivity $\kappa$ of the CoFe/[Cu/CoFe]$_{33}$ multilayer film at room temperature. The inset shows the $H$ dependence of the thermal conductance $G$ at the MgO/CoFe interface.

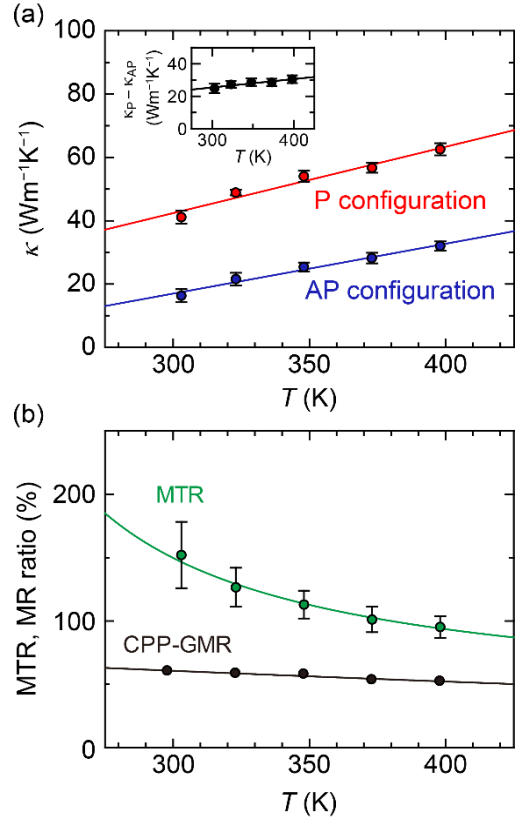

**Fig. 3.** Temperature dependence of magneto-thermal resistance effect. (a) Temperature $T$ dependence of $\kappa$ of the CoFe/[Cu/CoFe]$_{33}$ multilayer film for the P and AP configurations. The solid lines represent linear fits to the data. The inset shows the $T$ dependence of $\kappa_\mathrm{P} - \kappa_\mathrm{AP}$, where $\kappa_\mathrm{P(AP)}$ is the thermal conductivity for the P (AP) configuration. (b) $T$ dependence of the MTR ratio and the magnetoresistance (MR) ratio for current-perpendicular-to-plane giant magnetoresistance (CPP-GMR) in the Cu/CoFe multilayer films. The green solid curve is obtained from the solid lines in (a). The details for the CPP-GMR measurements are shown in Sec. 4 in supplementary material.

Supplementary Material

# Above-room-temperature giant thermal conductivity switching in spintronic multilayer

Hiroyasu Nakayama,[1,a)] Bin Xu,[2] Sotaro Iwamoto,[2] Kaoru Yamamoto,[1] Ryo Iguchi,[1] Asuka Miura,[1] Takamasa Hirai,[1] Yoshio Miura,[1] Yuya Sakuraba,[1,3] Junichiro Shiomi,[2,a)] and Ken-ichi Uchida[1,2,4,5,a)]

[1]National Institute for Materials Science, Tsukuba 305-0047, Japan.
[2]Department of Mechanical Engineering, The University of Tokyo, Tokyo 113-8656, Japan.
[3]PRESTO, Japan Science and Technology Agency, Saitama 332-0012, Japan.
[4]Institute for Materials Research, Tohoku University, Sendai 980-8577, Japan.
[5]Center for Spintronics Research Network, Tohoku University, Sendai 980-8577, Japan.

[a)]Author to whom correspondence should be addressed: NAKAYAMA.Hiroyasu@nims.go.jp, shiomi@photon.t.u-tokyo.ac.jp, and UCHIDA.Kenichi@nims.go.jp

## Supplementary Section 1. TDTR measurements

The TDTR system used in this study consists of a Ti:sapphire laser with a pulse width of 140 fs and a repetition rate of 80 MHz. The laser light is split into a pump pulse train (wavelength 400 nm, $1/e^2$ radius 12 μm, and modulation frequency 11.05 MHz) and a probe pulse train (wavelength 800 nm and $1/e^2$ radius 11 μm). The pump pulse induces impulse responses at the sample surface, and the probe pulse detects the temperature change of the Al transducer on the top of the Cu/CoFe multilayer through thermoreflectance. The reflected probe beam was detected by a Si photodiode connected to a lock-in amplifier to obtain the signal consisting of the in-phase voltage ($V_{in}$) and out-of-phase voltage ($V_{out}$) at the frequency of the laser light modulation, where the TDTR signal is defined by $-V_{in}(t)/V_{out}(t)$.[1,2] The power of the pump and probe beams was adjusted so that the temperature rise of the sample due to steady-state heating does not exceed 2 K for all the measurements.

## Supplementary Section 2. Procedures of sample preparation

The sample system used for the TDTR measurements consists of an epitaxial CoFe(3.0)/[Cu(1.6)/CoFe(3.0)]$_{33}$ multilayer film formed on a single-crystalline MgO (001) substrate, where the order of each layer is described from bottom to top and the numbers in parentheses indicate the layer thicknesses in nanometers [Fig. 2(a)]. In order to obtain the metastable bcc-Cu layers as well as an antiferromagnetic coupling between the CoFe layers, the same film thicknesses of the Cu/CoFe multilayer and same deposition conditions as the previous study[3] were adopted. The deposition of the multilayer film was performed at room temperature by using an ultrahigh vacuum magnetron sputtering system, where a base pressure was less than $1 \times 10^{-7}$ Pa. Before the deposition of the bottom CoFe layer, the surface of the MgO substrate was etched by Ar-ion milling in the sputtering chamber to obtain the (001)-oriented epitaxial growth of the Cu/CoFe multilayer. A 5.0-nm-thick Al transducer layer was then deposited on the multilayer by magnetron sputtering without breaking the vacuum. Finally, the sample was annealed at 250°C in vacuum with applying an in-plane magnetic field of 3 kOe for 1 hour to improve the film quality. The thickness of the multilayer was confirmed by the transmission electron microscopy. The magnetic properties of the Cu/CoFe multilayer film were measured by a vibrating sample magnetometry at room temperature. We also confirmed that the MR ratio for CIP-GMR of the CoFe/[Cu/CoFe]$_{33}$ multilayer film is 76%, which is comparable to the value reported in Supplementary Ref. 3, guaranteeing the same quality of our film as the Cu/CoFe multilayer used in the previous work.

## Supplementary Section 3. Heat conduction models for Cu/CoFe multilayer

To obtain sufficient sensitivity of the cross-plane thermal conductivity $\kappa$ of the Cu/CoFe multilayer and the interfacial thermal conductance $G$ between the MgO substrate and the bottom CoFe layer at the same time in the time-domain thermoreflectance (TDTR) method, the thickness of the transducer layer (Al) was set to 5.0 nm [Fig. S1(a)], while the thickness is typically set to ~100 nm. Since the transducer layer of the present system is too thin to absorb the pump laser energy completely, the laser energy is partially absorbed by the Cu/CoFe multilayer, making the initial temperature distribution different from the standard case using a thick transducer layer. In a standard TDTR method, owing to the high thermal conductivity of the transducer layer (~150 Wm$^{-1}$K$^{-1}$ for Al), the temperature distribution inside the transducer layer becomes uniform in a short time (< 200 ps), and thus, the initial temperature distribution has negligible influence to the temperature decay profile after 200 ps. Therefore, the heat conduction model (> 200 ps) can be simplified to a "one-directional heat conduction model", where the pump-laser-induced heating occurs directly on the sample surface and transports in one direction [Fig. S2(a)]. Since the initial temperature distribution in the present experimental system is different from the standard case, we need to verify whether it is reasonable to use the conventional one-directional heat conduction TDTR model in the present system.

A previous TDTR study on a Co(3.0)/[Cu(1.0)/Co(3.0)]$_{39}$/Ru(2.0) multilayer film on a MgO (001) single-crystalline substrate has discussed this issue in details.[1] To simplify the analysis of the influence of the initial temperature distribution on the time-domain temperature response, the authors of Supplementary Ref. 1 introduced a "bidirectional heat conduction model" [Fig. S2(b)]. In this model, they regarded the pump-laser-induced heating occurs at an artificial interface with a particular depth $\Delta d$ from the top surface of the sample based on the superposition principle.[1] By varying the depth of this artificial interface, the influence of the initial temperature distribution was studied. They found that the temperature decay curve in their system is insensitive to $\Delta d$, *i.e.*, initial temperature distribution, indicating the rationality of applying the typical one-directional heat conduction TDTR model to the system. Herein, we use the same bidirectional heat conduction model to analyze the influence of the initial temperature distribution on the temperature decay profile in the present Cu/CoFe multilayer. By changing $\Delta d$ as 0, 5, and 20 nm, the calculated temperature decay curves show significant differences for short delay time (~10 ps) but minor differences for delay time longer than 200 ps [Fig. S2(c)], which agrees well with the previous study. Since the laser penetration depth of the sample and the resultant $\Delta d$ value are no more than 20 nm, it is reasonable to use the standard one-directional TDTR model to solve the heat conduction in our Cu/CoFe multilayer film with a thin transducer layer.

To obtain $\kappa$ and $G$ through the fitting of the experimentally measured TDTR signal to the theoretical curve predicted by the heat conduction model, we applied a two-layer structure. The first layer of this two-layer structure consists of the CoFe/[Cu/CoFe]$_{33}$ multilayer and the Al transducer, which is regarded as a homogeneous medium to evaluate its effective thermal conductivity. We set the thickness of the metallic film to 159.8 nm, which is the total thickness of the CoFe/[Cu/CoFe]$_{33}$ multilayer film (154.8 nm) and the Al transducer (5.0 nm). We note that the contribution of the Al transducer in the effective thermal conductivity is negligibly small because of the small thickness and high thermal conductivity of the Al layer. The second layer of this two-layer structure is the MgO substrate. In the analysis, we used the specific heat capacities of the metallic film and MgO and the thermal conductivity of MgO [Fig. S3]. The specific heat capacity of the metallic film was calculated to be the weighted average of the bulk values[4] by considering the thickness and the composition of each layer. Then, the remaining unknown parameters in the model are the effective thermal conductivity $\kappa$ of the first layer and the interfacial thermal conductance $G$ between the first and second layers.

By fitting the TDTR signals with the model, we obtained the $\kappa$ and $G$ values. The fitting was performed for times later than 200 ps from the time of the impulse, to neglect the effect of electron-phonon relaxation[5] and the influence of the initial temperature distribution due to the thin transducer layer as discussed above. The obtained $G$ at room temperature is approximately 200 MWm$^{-2}$K$^{-1}$, which is comparable to that at MgO/metal interfaces reported in the previous works.[1,6]

## Supplementary Section 4. Current-perpendicular-to-plane giant magnetoresistance effect

To compare the magnetization configuration dependence of the cross-plane thermal conductivity with that of the electrical conductivity, we measured the current-perpendicular-to-plane (CPP)-giant magnetoresistance (GMR) effect by patterning the Cu/CoFe multilayer film into a pillar structure [Fig. S4(a)]. However, since it is difficult to construct a pillar of the CoFe/[Cu/CoFe]$_N$ multilayer films with large Cu/CoFe bilayer number $N$, we used the multilayer films with $N$ = 1, 3, 5, and 7 and investigated the $N$ dependence of the magnetoresistance (MR) ratio for CPP-GMR. The layer structure of the CPP-GMR device is Cr(10.0)/Ag(100.0)/CoFe(3.0)/[Cu(1.6)/CoFe(3.0)]$_N$/Ru(8.0) from bottom to top, where the thicknesses of the CoFe and Cu layers are the same as those of the sample used for the TDTR measurements. The CPP-GMR device was also formed on a MgO (001) single-crystalline substrate. Before the deposition of the Cr/Ag layer, the MgO substrate was flashed at 500°C for 30 minutes. Then, the films were deposited at room temperature by magnetron sputtering, where a base pressure was less than $1 \times 10^{-7}$ Pa. The Cr/Ag layers were annealed in situ at 300°C for 30 minutes to reduce the surface roughness of the Cr/Ag layers,[7] followed by the deposition of the CoFe/[Cu/CoFe]$_N$ multilayer in the same condition as the sample used for the TDTR measurements. The Cr/Ag layers act as a buffer layer for the multilayer as well as a bottom electrode, where atomically flat interfaces and highly oriented epitaxial growth on the Cr/Ag buffer layers are ensured by previous studies on CPP-GMR.[3,7] The CoFe/[Cu/CoFe]$_N$ multilayer was then patterned into a circular shape with diameters of 200 to 400 nm by means of electron beam lithography and Ar-ion milling. The cross-sectional area $A$ of the pillar was confirmed by scanning electron microscopy. We confirmed that the CPP-GMR output (resistance change-area product) is kept almost constant with changing $A$, which is consistent with the previous report.[8] Finally, the Ta(2.0)/Au(150.0) top electrode was deposited at room temperature by magnetron sputtering. The electrical resistance $R$ between the top and bottom electrodes was measured by a DC four probe method with applying an in-plane magnetic field $H$, where the $A$-dependent parasitic resistance is excluded by measuring the pillar-size dependence of the resistance.

    Figure S4(b) shows the $H$ dependence of $R$ of the CoFe/[Cu/CoFe]$_N$ device with $N$ = 1 at room temperature. We observed a clear GMR signal in this system, which shows the P (AP) configuration when $|H|$ > 50 Oe ($|H|$ < 25 Oe). The MR ratio for the CoFe/[Cu/CoFe]$_1$ device was observed to be ~5% in the CPP configuration, comparable to the value reported in a similar structure.[3] However, it is known that the electrical resistance of a sub-micron-scale pillar structure includes the parasitic resistance $R_\text{par}$, *e.g.*, interfacial resistance between the top electrode and capping layer of the pillar structure, in addition to the intrinsic CPP-GMR contribution; $R_{par}$ causes a significant decrease in the MR ratio for small $N$. Therefore, it is necessary to investigate the $N$ dependence of the MR ratio to exclude the $R_{par}$ contribution. Since the resistance change due to CPP-GMR, $R_{AP} - R_P$, in spintronic multilayers linearly increases with increasing $N$[7], the intrinsic MR ratio can be extracted in the limit of large $N$, where $R_{par} \ll R_P, R_{AP}$. We thus performed the same experiments using the CoFe/[Cu/CoFe]$_N$ devices with $N$ = 1, 3, 5, and 7 [Figs. S4(c)–S4(e)]. By extrapolating the $N$ dependence of the MR ratio and taking the $N \to \infty$ limit, we determined the intrinsic MR ratio MR$_{sat}$ for our CoFe/[Cu/CoFe]$_N$ multilayer film to be 60.4% at room temperature [see the inset of Fig. S4(e)], which is slightly smaller than the MR ratio for CIP-GMR of the Cu/CoFe multilayer films.[3] We repeated the same CPP-GMR measurements with changing the temperature $T$ and confirmed that MR$_{sat}$ monotonically decreases with increasing $T$ [Fig. 3(b)].

## Supplementary Section 5. First-principles calculations

In this section, we show the results of the first-principles calculations for the magneto-thermal resistance (MTR) effect. Here, we focus on spin-dependent ballistic electron transport at nonmagnetic/ferromagnetic metal interfaces, characterized by the electron transmittance (note that the calculations shown below do not include diffusive transport properties in bulk). We analyzed the transmission at Co/Cu/Co and CoFe/Cu/CoFe junctions and compared the MTR effect between these structures. We prepared an fcc-Co (9 ML)/fcc-Cu (7 ML)/fcc-Co (9 ML) (001) tetragonal supercell [Fig. S5(a)] and B2-CoFe (11 ML)/bcc-Cu (13 ML)/B2-CoFe (11 ML) (001) tetragonal supercells with Fe and Co terminations [Figs. S5(b) and S5(c)], where ML in the parentheses represents the number of monolayers. The in-plane lattice parameters are fixed to the lattice constant of the Cu spacer: $3.62/\sqrt{2}$ Å for the fcc-Cu spacer and 2.86 Å for the bcc-Cu spacer. The structural optimization for each supercell was carried out using the density-functional theory combined with the generalized gradient approximation for the exchange-correlation energy, which was implemented in Vienna Ab initio Simulation Package.[9]

The transport calculations were performed with Atomistix ToolKit (ATK)[10] by means of the linear combination of atomic orbitals and the pseudo-potential method. In the transport calculations, the bulk Co (CoFe) was attached to both sides of Co (CoFe)/Cu/Co (CoFe) supercell as semi-infinite electrodes in the Landauer formula. The effective potential for the structure was obtained from self-consistent-field calculations under an open boundary condition with a $7 \times 7 \times 50$ $k$-point mesh. Such a large number of $k_z$ points are required to minimize the mismatch of the Fermi energy between the electrode and the supercell. The transmittance was obtained by means of the non-equilibrium Green's function implemented in ATK. In the calculation for the CoFe/Cu/CoFe structure, we took the average of the transmittance of the structures with the Fe and Co terminations. Since our system was repeated periodically in the $x$-$y$ plane, propagating states were assigned an in-plane wave vector $\mathbf{k}_\parallel = (k_x, k_y)$ index. We set the in-plain $k$ points as $150 \times 150$ for the transport calculation. The electrical conductance $G_E$ and thermal conductance were obtained from the transmittance by means of the linear-response theory. The thermal conductance calculated here does not include the correction term due to the thermoelectric effect $G_E S_2 T$ with $S$ being the Seebeck coefficient, because we experimentally confirmed that the correction term is negligibly small.

Figures S5(d) and S5(e) shows the calculated transmittance for the Co/Cu/Co and CoFe/Cu/CoFe junctions as a function of the energy difference from the Fermi energy $E - E_F$. In comparison with the Co/Cu/Co junction, the CoFe/Cu/CoFe junction exhibits smaller transmittance around the Fermi energy in the AP configuration, while both the junctions exhibit comparable transmittance in the P configuration. The large difference in the transmittance between the P and AP configurations for the CoFe/Cu/CoFe junction originates from the high spin polarization of transmitted electrons [see the insets of Figs. S5(d) and S5(e)]. From the energy dependence of the transmittance, we calculated the temperature dependence of the MTR and MR ratios [see Figs. S5(f) and S5(g)]. Both the calculated MTR and MR ratios for the CoFe/Cu/CoFe junction are several times larger than those for the Co/Cu/Co junction over a wide temperature range owing to the difference in the transmittance in the AP configuration. Although we cannot compare the calculated values with the experimental results quantitatively since we focus only on the interfacial effect, our calculation clearly shows that the CoFe/Cu/CoFe junction is a suitable system for the MTR and GMR effects. We also found that, in the CoFe/Cu/CoFe junction, the calculated MTR ratio is greater than the MR ratio for $T > 540$ K and $T < 170$ K even when only the electronic thermal conductivity is considered [Fig. S5(g)]. This behavior is attributed to the facts that the transmittance shows a steep change near $E = E_F$ and that the integrand in the formulation of the thermal conductance has a different $E$- and $T$-dependent factor from that of the electrical conductance.[11] However, the calculated difference between the MTR and MR ratios is too small to explain the experimental results in Fig. 3(b), where the MTR ratio for the Cu/CoFe multilayer film was observed to be twice as large as the MR ratio.

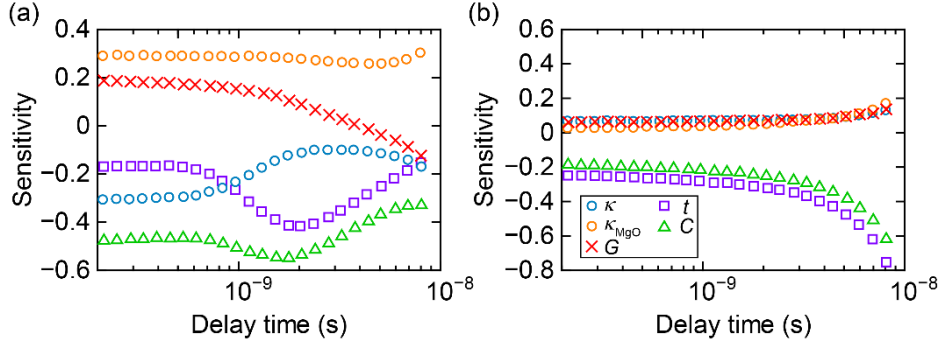

**FIG. S1.** Sensitivity calculations for TDTR measurements. (a), (b) Sensitivity analysis for the CoFe/[Cu/CoFe]$_{33}$ multilayer films on a MgO (001) substrate with 5-nm-thick (a) and 90-nm-thick (b) Al transducer layers. $\kappa$, $\kappa_{MgO}$, $G$, $t$, and $C$ are the effective thermal conductivity of the multilayer film, thermal conductivity of the MgO substrate, interfacial thermal conductance between MgO and CoFe, thickness of the multilayer, and specific heat capacity of the multilayer, respectively.

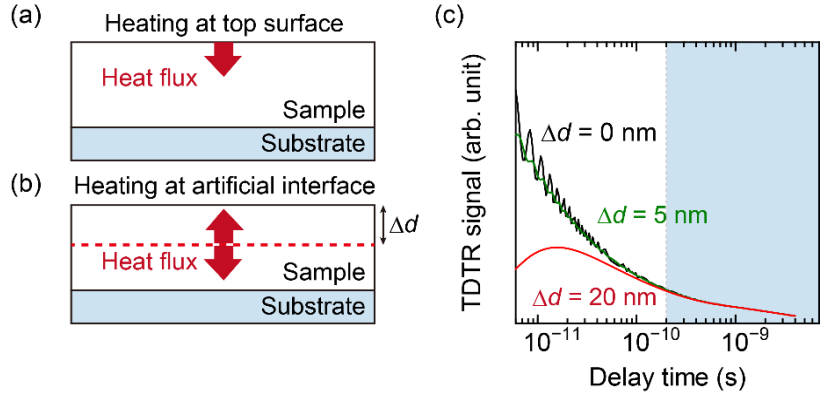

**FIG. S2.** Heat conduction models for Cu/CoFe multilayer. (a), (b) Schematics of one-directional (a) and bidirectional (b) heat conduction models. (c) Calculated temperature decay profile based on the bidirectional heat conduction model, which corresponds to temporal response of TDTR signals, for various depths of the artificial interface $\Delta d$ of pump-laser-induced heating. The region in light blue is used for the TDTR analyses.

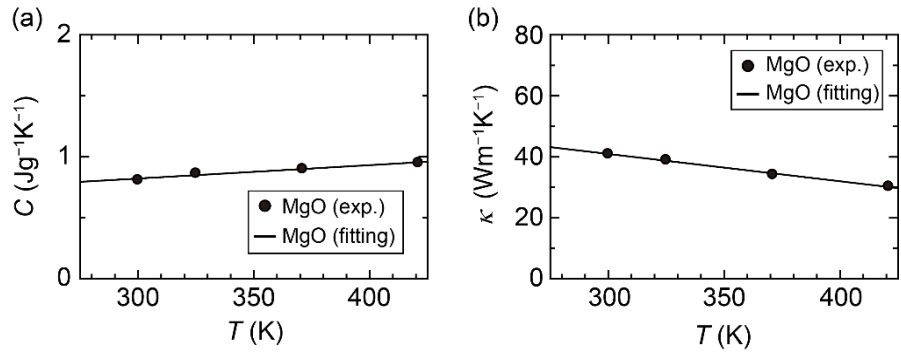

**FIG. S3.** Thermal properties of MgO substrate. (a) Temperature $T$ dependence of the specific heat capacity $C$ of a MgO (001) single-crystalline substrate, measured by the differential scanning calorimetry. (b) $T$ dependence of the thermal conductivity $\kappa$ of the MgO substrate, determined through the thermal diffusivity measured by the laser flash method, the $C$ values in (a), and density of MgO (= 3.58 g cm$^{-3}$) measured by the Archimedes method. The solid lines are linear fits to the data. The $C$ and $\kappa$ values used for the TDTR analyses are determined by the linear interpolation.

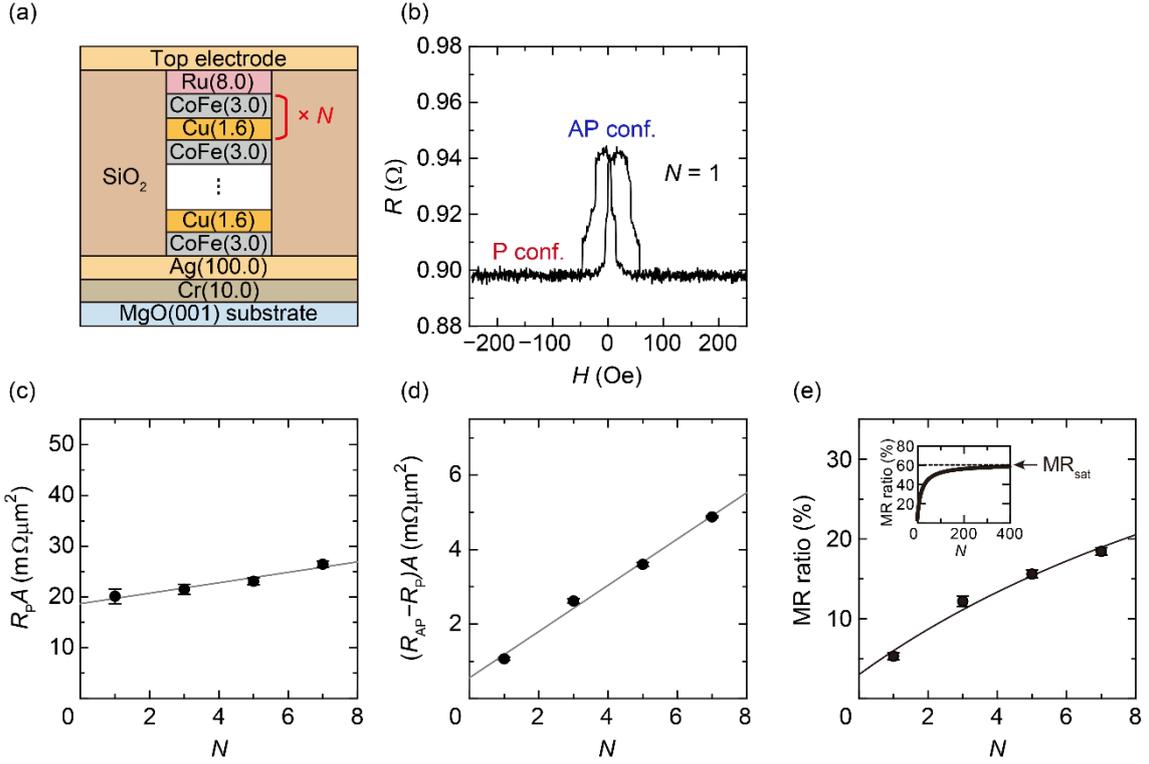

**FIG. S4.** Current-perpendicular-to-plane giant magnetoresistance. (a) Schematic of the CoFe/[Cu/CoFe]$_N$ pillar device used for the CPP-GMR measurements, where the Cu/CoFe bilayer number is $N = 1, 3, 5$, and 7. The numbers in parentheses indicate the layer thicknesses in nanometers. (b) The in-plane magnetic field $H$ dependence of the cross-plane electrical resistance $R$ of the CoFe/[Cu/CoFe]$_1$ device at room temperature, where the diameter of the pillar is 200 nm. (c), (d) $N$ dependence of the resistance area product $R_P A$ and resistance-change area product $(R_{AP} - R_P)A$ for the CoFe/[Cu/CoFe]$_N$ devices. $R_{P(AP)}$ is the electrical resistance for the P (AP) configuration and $A$ is the cross-sectional area of the pillar. The gray solid lines are linear fits to the data. Both $R_P A$ and $(R_{AP} - R_P)A$ linearly increase with increasing $N$, indicating that the quality of all the Cu/CoFe interfaces is similar in the present devices. (e) $N$ dependence of the MR ratio $(R_{AP} - R_P)/R_P$. Each data point was obtained from an averaged resistance value for 10 or more devices. The black solid curve is obtained from the data in (c) and (d). The inset shows the $N$ dependence of the extrapolated MR ratio. The saturated MR ratio $MR_{sat}$ shows the intrinsic CPP-GMR performance.

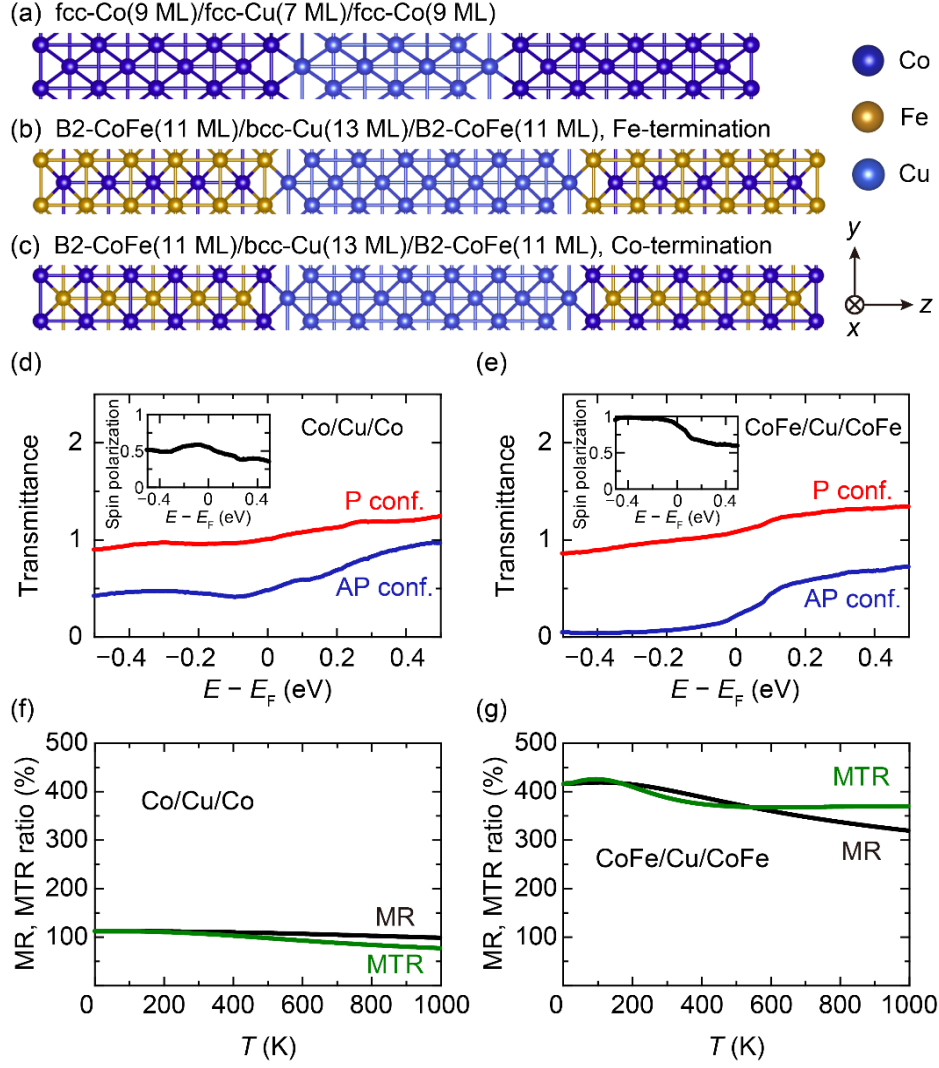

**FIG. S5.** First-principles calculations. (a)–(c) Supercells of the fcc-Co/fcc-Cu/fcc-Co (001) junction (a), B2-CoFe/bcc-Cu/B2-CoFe (001) junction with the Fe termination (b), and B2-CoFe/bcc-Cu/B2-CoFe (001) junction with the Co termination (c). (d), (e) Transmittance across the Co/Cu/Co (d) and CoFe/Cu/CoFe (e) junctions as a function of the energy difference from the Fermi energy $E - E_F$. The transmittance can be > 1 because it is the sum of the transmittance from bands contributing to the transport. The insets show the $E - E_F$ dependence of the spin polarization for the transmittance across the Co/Cu/Co and CoFe/Cu/CoFe junctions, which is defined by $(\tau_{up} - \tau_{down})/(\tau_{up} + \tau_{down})$ with $\tau_{up(down)}$ being the transmittance of up-spin (down-spin) electrons. (f), (g) $T$ dependence of the calculated MTR and MR ratios for the Co/Cu/Co (f) and CoFe/Cu/CoFe (g) junctions.